\documentclass[10pt,journal,doublecolumn]{IEEEtran}
\usepackage{cite}
\usepackage{amsmath,amssymb,placeins}
\usepackage{algorithmic}
\usepackage{graphicx}
\usepackage{textcomp}
\usepackage[mathscr]{euscript}
\usepackage{lipsum}
\usepackage{cuted}
\usepackage{mathtools}
\usepackage{multicol}
\usepackage{array}
\usepackage{makecell}
\usepackage{diagbox}
\usepackage[symbol]{footmisc}
\usepackage{caption}
\usepackage{tabularx}

\usepackage[mathscr]{euscript}
\usepackage[usenames]{color}
\usepackage{amsfonts}
\usepackage{latexsym}
\usepackage{subfigure}
\usepackage[format=hang]{subfig}
\usepackage{caption}
\usepackage{floatrow}
\usepackage{multirow}
\usepackage{epstopdf}
\newtheorem{theorem}{{{\textit{Theorem}}}}

\newtheorem{lemma}{{{\textit{Lemma}}}}

\newtheorem{definition}{{{\textit{Definition}}}}

\newtheorem{remark}{{{\textit{Remark}}}}

\newtheorem{example}{{{\textit{Example}}}}

\def\BibTeX{{\rm B\kern-.05em{\sc i\kern-.025em b}\kern-.08em
		T\kern-.1667em\lower.7ex\hbox{E}\kern-.125emX}}
		
		\def\@fnsymbol#1{\ensuremath{\ifcase#1\or *\or \dagger\or \ddagger\or
   \mathsection\or \mathparagraph\or \|\or **\or \dagger\dagger
   \or \ddagger\ddagger \else\@ctrerr\fi}}

\begin{document}
\title{A Direct Construction of Cross Z-Complementary Sequence Sets with Large Set Size}
\author{Praveen Kumar, Sudhan Majhi, ~\IEEEmembership{Senior Member,~IEEE,}
       Subhabrata Paul  
  \thanks{Praveen Kumar and Subhabrata Paul are with the Department of Mathematics, IIT Patna, Bihta, Patna, 801106, Bihar, India (e-mail: praveen\_2021ma03@iitp.ac.in; subhabrata@iitp.ac.in). }   
  \thanks{Sudhan Majhi is with the Department of Electrical Communication Engineering (IISC), Indian Institute of Science, Bangalore, CV Raman Rd, 560012, Karnataka, India (email:smajhi@iisc.ac.in) }  }  

\IEEEpeerreviewmaketitle
\maketitle
\begin{abstract}
This letter presents a direct construction of cross Z-complementary sequence sets (CZCSSs), whose aperiodic correlation sums exhibit zero correlation zones at both the front-end and tail-end shifts. 
 CZCSS can be regarded as an extension of the symmetrical Z-complementary code set (SZCCS). The available construction of SZCCS has a limitation on the set size, with a maximum set size of 8. The proposed generalized Boolean function based construction can generate CZCSS of length in the form of non-power-of-two with variable set size $2^{n+1}$, where each code has $2^{n+1}$ constituent sequences. The proposed construction also yields cross Z-complementary pairs and cross Z-complementary sets with a large number of constituent sequences compared to the existing work.
\end{abstract}
\section{Introduction}\label{sec:intro}
\IEEEPARstart{T}{he} idea of Z-complementary pairs (ZCPs) was introduced by Fan \textit{et al.} \cite{fan2007}. Within a certain zone, known as the zero-correlation zone (ZCZ), the aperiodic auto-correlation sum (AACS) of the two sequences in a ZCP is zero. When this ZCZ width $Z$ is equal to the sequence length $N$, ZCP becomes a Golay complementary pair (GCP). Unlike GCPs, ZCPs are available in arbitrary lengths with a variety of ZCZ widths \cite{zcp1,zcp2,zcp3,zcp4,praveen1}. 

Liu \textit{et al.} examined the training sequence design for the spatial modulation (SM) system in \cite{czcp2}, leading to the development of the cross Z-complementary pair (CZCP). CZCPs have both front-end and tail-end ZCZ for AACS and tail-end ZCZ for aperiodic cross-correlation sum (ACCS). 
 The cross-ZCZ (CZCZ) ratio of a CZCP is defined as the ratio of ZCZ width $Z$ and maximum possible ZCZ width $Z_{max}$. The maximum possible ZCZ width for a CZCP of length $N$ is $N/2$, i.e., $Z_{max} =N/2$. A CZCP is called perfect if its CZCZ ratio takes the value equal to $1$.
 Several constructions of CZCPs with different lengths and CZCZ ratios can be found in \cite{czcp1,czcp3,czcp4,czcp5,czcp6,czcp7}. Recently, the CZCPs are extended to cross Z-complementary set (CZCS) \cite{czcs}.  

The idea of ZCPs introduced in \cite{fan2007} was generalized to Z-complementary code set (ZCCS)
by Feng \textit{et al.} in \cite{zccs1}.
ZCCSs only consider the front-end ZCZ of the AACSs and ACCSs \cite{pmz,zccs2,zccs3,gobinda}. 
Recently,  the idea of ZCCS is extended to symmetrical-ZCCS (SZCCS) which exhibits both the front-end and tail-end
ZCZ properties \cite{szccs1}. The authors in \cite{szccs1}, have presented a generalized Boolean functions (GBFs) based construction of $(8,8,2^m,2^{m-2}-1)$-SZCCS. In practice, a front-end ZCZ and a tail-end ZCZ have particular interest for mitigating interference with small and large delays, respectively. The cross Z-complementary
sequence set (CZCSS) is a generalization of SZCCS, which satisfies all the properties of SZCCS and have some extra correlation properties satisfying for tail-end ZCZ. 
SZCCSs with larger set size are used in designing training sequences for broadband generalized spatial modulation (GSM) systems over frequency-selective channels \cite{szccs1}. 

In this letter, we proposed a GBFs based construction of CZCSSs of non-power-of-two length whose set size is the same as the number of constituent sequences. To generate the desired CZCSSs, first a new CZCPs are constructed, then by using the proposed CZCPs and CZCPs of \cite{czcp1}, CZCSSs with large set size are constructed. To the best of the authors' knowledge, the proposed construction of CZCSSs is not reported in the existing literature. Since CZCS and SZCCS are special cases of the proposed CZCSS, so this letter also facilitates these constructions with large flock size and set size, respectively.            

The rest of the letter is arranged in the following manner. Section II covers the fundamental notations and definitions. Section III discusses the proposed CZCSS construction based on GBF. Section IV has final observations.
\section{Notations and Definitions}
The basic notations, definitions, and earlier known findings that are necessary for the proposed construction are explained in this section.
\begin{definition}\label{def1} Let $\mathbf{c_1}$ $=(c_{10},c_{11}, \ldots, c_{1L-1})$ and $\mathbf{c_2}$ $=(c_{20},c_{21}, \ldots, c_{2L-1})$ be two $L$ length sequences over $\mathbb{Z}_{q}$. At a shift $\tau$, the aperiodic cross-correlation function (ACCF) is defined by
\begin{equation}\label{eqn 1}
C\left({\mathbf{c}, \mathbf{d}}\right)(\tau)=\begin{cases}
\sum_{i=0}^{L-1-\tau} \omega^{c_{1i}-c_{2i+\tau}}, & 0 \leq \tau \leq L-1, \\
\sum_{i=0}^{L-1+\tau} \omega^{c_{1i-\tau}-c_{2i}}, & -L+1 \leq \tau \leq-1, \\
0, & |\tau| \geq L,
\end{cases}
\end{equation}
 where $q~(\geq 2)$ is an integer and $\omega=\exp(2\pi\sqrt{-1}/q)$. When $\mathbf{c_1}=\mathbf{c_2},~ A({\mathbf{c_1}, \mathbf{c_2}})(\tau)$ is called aperiodic auto-correlation function (AACF) of $\mathbf{c_1}$ and it is denoted by $A(\mathbf{c_1})(\tau)$.
 \end{definition}
\begin{definition}\label{def2} A ZCP with ZCZ $Z$ is a pair of sequences $\mathbf{c}$ and $\mathbf{d}$ of length $N$ whose AACS is zero for all non-zero shifts inside a zone, i.e.,
\begin{equation}\label{eqn2}
A({\mathbf{c}})(\tau)+A({\mathbf{d}})(\tau)=0, \text{ for all }  0 <\tau < Z.
\end{equation}
When $Z=N$, the pair ($\mathbf{c},\mathbf{d}$) is known as a GCP.
\end{definition}
\begin{definition}\label{def3} 
 For an integer $Z$ and $N$, let $\mathcal{U}_{1}=\{1,2, \ldots, Z\}$ and $\mathcal{U}_{2}=\{N-Z, N-Z+1, \ldots, N-1\}$. Then a length $N$ sequence pair $(\mathbf{c}, \mathbf{d})$ is called an $(N, Z)$-CZCP if it satisfies the following two conditions:
\begin{equation}\label{eqn3}
 \begin{aligned}
&{C1}: A({\mathbf{c}})(\tau)+A({\mathbf{d}})(\tau)=0 \text {, for all }|\tau| \in \mathcal{U}_{1} \cup \mathcal{U}_{2};\\&
{C2}: C({\mathbf{c}, \mathbf{d}})(\tau)+C({\mathbf{d}, \mathbf{c}})(\tau)=0, \text{ for all } |\tau| \in \mathcal{U}_{2} .
\end{aligned}  
\end{equation}
\end{definition}
\begin{definition}\label{def4}
  Consider a set  $\mathcal{C}=\left\{{C^{0}}, {C^{1}}, \hdots, {C^{K-1}}\right\}$, where each set ${C^{p}}$ consists of $M$ sequences, i.e., ${C^{p}}=\left\{\mathbf{c}_{0}^{p}, \mathbf{c}_{1}^{p}, \hdots, \mathbf{c}_{M-1}^{p}\right\}$, and length of each sequence $\mathbf{c}_{l}^{p}$ is $N$, where $0 \leq p \leq K-1$ and $0 \leq l \leq M-1$. The set $\mathcal{C}$ is called a CZCSS, denoted by $(K,M,N,Z)$-CZCSS, if for $\mathcal{U}_{1}=\{1,2, \cdots, Z\}$ and $\mathcal{U}_{2}=\{N-Z, N-$ $Z+1, \cdots, N-1\}$ with $Z \leq N,$ it satisfies the following properties
\begin{equation}\label{eqn4}
    \begin{aligned}
      &P1: \sum_{i=0}^{M-1} A(\mathbf{c}_i^p)(\tau)=0, \quad \text { for all }|\tau| \in \mathcal{U}_{1} \cup \mathcal{U}_{2};  \\&
      P 2: \sum_{i=0}^{M-1} C(\mathbf{c}_i^p, \mathbf{c}_{i+1}^p)(\tau)=0, \quad \text{ for all } |\tau| \in \mathcal{U}_{2};\\&
      P 3: \sum_{i=0}^{M-1} C(\mathbf{c}_i^p, \mathbf{c}_i^{p'})(\tau)=0, \quad \text{ for all } |\tau| \in\{0\} \cup \mathcal{U}_{1} \cup \mathcal{U}_{2};\\&
      P 4: \sum_{i=0}^{M-1} C(\mathbf{c}_i^p, \mathbf{c}_{i+1}^{p'})(\tau)=0, \quad \text{ for all }|\tau| \in \mathcal{U}_{2};
    \end{aligned}
\end{equation}
where $\mathbf{c}_{M}^{p}=\mathbf{c}_{0}^{p}, \mathbf{c}_{M}^{p'}=\mathbf{c}_{0}^{p'}$ and $p \neq p'$.
\end{definition}
The complex-valued sequence corresponding to a GBF $f:\{0,1\}^m \rightarrow \mathbb{Z}_q$ of $m$ variables $x_0,x_1,\hdots,x_{m-1}$ is expressed as \cite{czcp1}
\begin{equation}\label{eqn5}
 \Psi(f)=\left(\omega^{f_0}, \omega^{f_1}, \hdots, \omega^{f_{2^m-1}}\right),
\end{equation}
where $f_i=f(s_{i,0},s_{i,1},\hdots,s_{i,m-1})$, $\omega=\exp\left(2\pi\sqrt{-1}/q\right)$, and $\mathbf{s}_i=(s_{i,0},s_{i,1},\hdots,s_{i,m-1})$ is the 
binary vector representation of the integer $i$.
Corresponding to a GBF $f$ with $m$ variables the sequence $\Psi(f)$ is of length $2^{m} .$ This letter focuses on $q$-ary CZCSSs of length in the form of non-power-of-two, where $q$ is a positive even integer throughout the remainder of this letter. The truncated complex-valued sequence $\Psi_{L}(f)$ corresponding to the GBF $f$ is defined here by eliminating the first and final $L$ components of the sequence $\Psi(f)$.


\begin{lemma}[\cite{wuyu}]\label{lem1} For any permutation $\pi$ of $\{0,1,2,\ldots ,m-1\}$ and constants $c, c_{i} \in \mathbb{Z}_{q}$, let $g:\{0,1\}^{m} \rightarrow \mathbb{Z}_{q}$ be a GBF defined by
\begin{equation}\label{eqn6}
g=\frac{q}{2} \sum_{i=0}^{m-2} x_{\pi(i)} x_{\pi(i+1)}+\sum_{i=0}^{m-1} c_{i} x_{i}+c.
\end{equation}
Then for any arbitrary constant $c'\in \mathbb{Z}_{q}$, the sequence pair
$(\mathbf{a}, \mathbf{b})$=$\big(\Psi(g)$,$ \Psi\left(g+\frac{q}{2} x_{\pi(0)}+c'\right)\big)$ is a GCP and the pair $(\mathbf{c}, \mathbf{d}) =$
$\big(\Psi\left(g+\frac{q}{2} x_{\pi(m-1)}\right)$,$ \Psi(g+\frac{q}{2}\left(x_{\pi(0)}+x_{\pi(m-1)}\right)+c'\big)$ is a complementary mate of $(\mathbf{a}, \mathbf{b})$ \cite{wuyu}.
\end{lemma}
\section{Proposed Construction}
In this section, a direct construction of CZCSS of length in the form of non-power-of-two based on GBF is presented. To construct the CZCSS, first a new CZCP is constructed and using this proposed and existing CZCP \cite{czcp1}, CZCSSs are constructed.

For any integer $m\geq 4$, let $\pi$ be a permutation of $\{0,1,\hdots,m-3\}$. Let us define a GBF $G:\{0,1\}^{m}$ $\rightarrow$ $\mathbb{Z}_q$ as 
\begin{equation}\label{eqn7}
    G=\frac{q}{2}\left(\bar{x}_{m-1}x_{m-2} g+x_{m-1}\bar{x}_{m-2}  \left(g+x_{\pi(0)}+m-2\right)\right)+c,
\end{equation}
where $\bar{x}=1-x,c \in \mathbb{Z}_q$ is any constant and  $g:\{0,1\}^{m-2}$ $\rightarrow$ $\mathbb{Z}_q$ is a GBF defined as
\begin{equation}\label{eqn8}
    g=\sum_{\alpha=0}^{m-4} x_{\pi(\alpha)} x_{\pi(\alpha+1)},
\end{equation}
As \textit{Lemma} \ref{lem1}, $(g,g+x_{\pi(m-3)})$ is a GCP  and $(g+x_{\pi(0)}+m-2,g+x_{\pi(0)}+x_{\pi(m-3)}+m-2)$ is its complementary mate, the following result follows.
\begin{lemma}[\cite{czcp1}]\label{lem2}
Let $G_1=G+\frac{q}{2}(x_{m-1}x_{m-2})$ and $G_2=G+\frac{q}{2} x_{\pi(m-3)} \left(x_{m-1}+x_{m-2}\right)$, then the pair of sequences $(\mathbf{a}, \mathbf{b})=\left(\Psi_{2^{m-2}-1}(G_1), \Psi_{2^{m-2}-1}\left(G_2\right)\right)$ forms a $\left(2^{m-1}+2,2^{\pi(m-3)}+1\right)$-CZCP.
\end{lemma}
\begin{lemma}\label{lem3}
For GBF $g$ defined in (\ref{eqn8}) and for any shift $2^{m-2}-2^{\pi(m-3)} \leq \tau \leq 2^{m-2}-1$, the following cross-correlation property holds
\begin{equation}\label{eqn9}
 C\left(\mathbf{p},\mathbf{q}\right)(\tau) + C\left(\mathbf{u},\mathbf{v}\right)(\tau)  =0,
\end{equation}
where, for any $c_1\in \mathbb{Z}_q$, $\mathbf{p}=\Psi\left(\frac{q}{2} g\right)$, $\mathbf{q}=\Psi\big(\frac{q}{2}(g+x_{\pi(0)}+x_{\pi(m-3)}+c_1\big)$, $\mathbf{u}=\Psi\big(\frac{q}{2}(g+x_{\pi(m-3)})\big)$ and $\mathbf{v}=\Psi\big(\frac{q}{2}(g+x_{\pi(0)}+c_1)\big)$.
\end{lemma}
 \begin{IEEEproof}
 For a given integer $i$, set $j=i+\tau$ and let $(i_0,i_1,\hdots,i_{m-1})$ and $(j_0,j_1,\hdots,j_{m-1})$ be the binary
representation of $i$ and $j$, respectively. Then the value of ACCS is given by
\begin{equation}\label{eqn10}
\begin{aligned}
 &C\left(\mathbf{p},\mathbf{q}\right)(\tau) + C\left(\mathbf{u},\mathbf{v}\right)(\tau)
\\=&\sum_{i=0}^{N-\tau}\left((-1)^{g_i-g_j}\left((-1)^{-j_{\pi(0)}-j_{\pi(m-3)}}+(-1)^{i_{\pi(m-3)}-j_{\pi(0)}}\right)\right),
 \end{aligned}
\end{equation}
where $N=2^{m-2}-1.$ Since $2^{m-2}-2^{\pi(m-3)} \leq \tau \leq 2^{m-2}-1$, so $i_{\pi(m-3)}\neq j_{\pi(m-3)}$. Let if possible $i_{\pi(m-3)}= j_{\pi(m-3)}$, then the value of $\tau$ is calculated as
\begin{equation}\label{eqn11}
\begin{aligned}
    &\tau=j-i=(j_0-i_0)2^0+(j_1-i_1)2^1+\cdots\\&+(j_{\pi(m-3)}-i_{\pi(m-3)})2^{\pi(m-3}+\cdots+(j_{m-3}-i_{m-3})2^{m-3}\\&
    \leq 2^0+2^1+\cdots+2^{m-3}-2^{\pi(m-3)} \leq 2^{m-2}-2^{\pi(m-3)}-1,
    \end{aligned}
\end{equation}
i.e., $\tau < 2^{m-2}-2^{\pi(m-3)}$, which is a contradiction. So $i_{\pi(m-3)}\neq j_{\pi(m-3)}$ and the result follows from (\ref{eqn10}).
 \end{IEEEproof}
\begin{lemma}\label{lem4}
The sequence pair $(\mathbf{c}, \mathbf{d})=\left(\Psi_{2^{m-2}-1}(G_1+\frac{q}{2}x_{m-2}), \Psi_{2^{m-2}-1}\left(G_2+\frac{q}{2}x_{m-2}\right)\right)$ forms a $\left(2^{m-1}+2,2^{\pi(m-3)}+1\right)$-CZCP. Additionally, the CZCPs $(\mathbf{a}, \mathbf{b})$ and $(\mathbf{c}, \mathbf{d})$ satisfy the following two properties: 
\begin{equation}\label{eqn12}
 C(\mathbf{a},\mathbf{c})(\tau)+C(\mathbf{b},\mathbf{d})(\tau)=0, \text{ for all } \tau,
 \end{equation}
 and
 \begin{equation}\label{eqn13}
 \begin{aligned}
     C(\mathbf{a},\mathbf{d})(\tau)\!+\!C(\mathbf{b},\mathbf{c})(\tau)\!=\!0,  2^{m-1}\!-\!2^{\pi(m-3)}\! < \tau \leq 2^{m-1}\!+\!1.
     \end{aligned}
    \end{equation}
\end{lemma}
\begin{IEEEproof}
The proof of ($\mathbf{c},\mathbf{d}$) as a CZCP can be done similarly as the proof of \textit{Lemma} \ref{lem2} \cite{czcp1}. Next, we provide a brief proof of the property in (\ref{eqn12}). For $\tau=0$, the result follows immediately from \textit{Lemma} \ref{lem1}. Now, for different values of $\tau$, the proof is splitted into different cases.

\textit{Case I}: $0<\tau\leq 2^{m-2}$.
Using \textit{Lemma} \ref{lem1} the ACCS is given by
\begin{equation}\label{eqn14}
\begin{aligned}
   &C(\mathbf{a},\mathbf{c})(\tau)+ C(\mathbf{b},\mathbf{d})(\tau)\\=&-\omega^{\frac{q}{2}g\left(\mathbf{s}_{2^{m-2}+\tau-1}\right)}\left(1+\omega^{\frac{q}{2}\times s_{2^{m-2}+\tau-1,\pi(m-3)}}\right)\\&\!+\!\omega^{\frac{q}{2}\left(g+x_{\pi(0)}+m-2\right)\left(\mathbf{s}_{3\times 2^{m-2}-\tau}\right)}\left(1-\omega^{\frac{q}{2}\times s_{3\times 2^{m-2}-\tau,\pi(m-3)}}\! \right),
   \end{aligned}
\end{equation}
where $g$ is defined in (\ref{eqn8}.)\\
\textit{Subcase I}: $s_{2^{m-2}+\tau-1,\pi(m-3)}=1$ and $s_{3\times 2^{m-2}-\tau,\pi(m-3)}$$=$$0$. In this case, the sum in (\ref{eqn14}) is 0.\\
\textit{Subcase II}: $s_{2^{m-2}+\tau-1,\pi(m-3)}=0$ and $s_{3\times 2^{m-2}-\tau,\pi(m-3)}=1$. Since $2^{m-2}+\tau-1+3\times 2^{m-2}-\tau=2^m-1.$ So, the binary vector representation of $s_{2^{m-2}+\tau-1,\pi(m-3)}=0$ is complement of $s_{3\times 2^{m-2}-\tau,\pi(m-3)}.$ Hence $(g+x_{\pi(0)}+m-2)\left(\mathbf{s}_{3\times 2^{m-2}-\tau}\right)$$=$$g\left(\mathbf{s}_{2^{m-2}+\tau-1}\right)$, and sum in (\ref{eqn14}) is 0.

\textit{Case II}: $2^{m-2}\leq\tau < 2^{m-2}+2^{\pi(m-3)}$.
\begin{equation}\label{eqn15}
\begin{aligned}
   &C(\mathbf{a},\mathbf{c})(\tau)+ C(\mathbf{b},\mathbf{d})(\tau)\\=&\omega^{\frac{q}{2}\left(g+x_{\pi(0)}+m-2\right)\left(\mathbf{s}_{2^{m-2}+\tau-1}\right)}\left(1+\omega^{\frac{q}{2}\times s_{2^{m-2}+\tau-1,\pi(m-3)}}\right)\\&+\omega^{\frac{q}{2}g\left(\mathbf{s}_{3\times 2^{m-2}-\tau}\right)}\left(1-\omega^{\frac{q}{2}\times s_{3\times 2^{m-2}-\tau,\pi(m-3)}} \right).
   \end{aligned}
\end{equation}
Now, again considering two different \textit{subcases}, as in \textit{Case I} the sum in (\ref{eqn15}) becomes zero.

\textit{Case III}: $\tau >2^{m-2}+2^{\pi(m-3)}$.\\
The sum $C(\mathbf{a},\mathbf{c})(\tau)+ C(\mathbf{b},\mathbf{d})(\tau)$ take the same values as in (\ref{eqn15}). From \cite[Lemma 4]{czcp1}, the only possible value of $s_{2^{m-2}+\tau-1,\pi(m-3)}=1$ and $s_{3\times 2^{m-2}-\tau,\pi(m-3)}=0$, and hence the sum in (\ref{eqn15}) is zero.

So, from \textit{Case I- Case III}, the result in (\ref{eqn12}) follows. Next, the property of (\ref{eqn13}) will be proved.
Now, for $2^{m-1}-2^{\pi(m-3)} < \tau < 2^{m-1}+1$, the ACCS of sequence pairs $(\mathbf{a},\mathbf{d})$ and $(\mathbf{b},\mathbf{c})$ is given by
\begin{equation}\label{eqn16}
    \begin{aligned}
        &C(\mathbf{a},\mathbf{d})(\tau)+C(\mathbf{b},\mathbf{c})(\tau)=C\left(\mathbf{p},\mathbf{q}\right)(u)+C\left(\mathbf{u},\mathbf{v}\right)(u)\\&\!+\! \omega^{-\frac{q}{2}\left(g+x_{\pi(0)}+m-2\right)\!\left(\mathbf{s}_{2^{m-2}+\tau-1}\right)}\left(\omega^{-\frac{q}{2}\times {s}_{2^{m-2}+\tau-1,\pi(m-3)}}+1\! \right)\\&+
        \omega^{\frac{q}{2}g\left(\mathbf{s}_{3\times 2^{m-2}-\tau}\right)}\left(1-\omega^{\frac{q}{2}\times {s}_{3\times 2^{m-2}-\tau,\pi(m-3)}} \right),
    \end{aligned}
\end{equation}
where $u=\tau-2^{m-2}-1$. Since $2^{m-1}-2^{\pi(m-3)} < \tau < 2^{m-1}+1$, then $2^{m-2}-2^{\pi(m-3)}\leq u \leq 2^{m-2}-1$. So, from \textit{Lemma} \ref{lem3}, $C\left(\mathbf{p},\mathbf{q}\right)(u)+C\left(\mathbf{u},\mathbf{v}\right)(u)=0$, and from \cite[Lemma 4]{czcp1}, the only possible values of  ${s}_{2^{m-2}+\tau-1,\pi(m-3)}=1$ and ${s}_{3\times 2^{m-2}-\tau,\pi(m-3)}=0$, and hence the sum in (\ref{eqn16}) is zero. Now for $\tau=2^{m-1}+1$, the cross-correlation sum is given by
\begin{equation}\label{eqn17}
    \begin{aligned}
        &C(\mathbf{a},\mathbf{d})(\tau)+C(\mathbf{b},\mathbf{c})(\tau)=\omega^{q/2}+\omega^0=0.
        \end{aligned}
\end{equation}
Hence, the results follows from  (\ref{eqn16}) and (\ref{eqn17}).
\end{IEEEproof}
\begin{example}\label{ex1}
Let $m=5,q=4$ and $\pi(0)=1,\pi(1)=0,\pi(2)=2$ and $G:\{0,1\}^4 \rightarrow \mathbb{Z}_q$ be a GBF defined as 
\begin{equation}\label{eqn18}
    G=2\left(\bar{x}_4x_3g+x_4\bar{x}_3\left(g+x_1+3\right)\right),
\end{equation}
where $g$ is a GBF defined as $g=x_1x_0+x_0x_2$. Also let $G_1=G+2x_4x_3$ and $G_2=G+2x_2(x_4+x_3)$, then from \textit{Lemma} \ref{lem2}, the pair of sequences $(\mathbf{a}, \mathbf{b})=\left(\Psi_7(G_1), \Psi_7\left(G_2\right)\right)$ forms a $\left(18,5\right)$-CZCP. Also from \textit{Lemma} \ref{lem4}, the pair $(\mathbf{c}, \mathbf{d})=\left(\Psi_7(G_1+2x_3), \Psi_7\left(G_2+2x_3\right)\right)$ is a $\left(18,5\right)$-CZCP.
The sequence pairs are given below explicitly, where $i$ represents $\omega^i$.\\
$\mathbf{a}=( 0     0     0     0     2     0     2     0     0     2     2     0    2     2    0    0    0  2),$ 
$\mathbf{b}=(0     0     0     0     2     2     0     2     2    2     2    0   2     0    2    2    2 0)$. \\
 $\mathbf{c}=(0     2     2     2     0     2     0     2     2     2     2     0    2     2     0     0    0 0)$,
$\mathbf{d}=( 0     2     2     2     0     0     2     0     0     2     2     0    2     0    2    2
2    2)$. \\
Additionally the sequence pairs $(\mathbf{a}, \mathbf{b})$ and $(\mathbf{c}, \mathbf{d})$ satisfy the following two properties\\
$C(\mathbf{a},\mathbf{c})(\tau)+C(\mathbf{b},\mathbf{d})(\tau)=0$, for all $\tau$.\\
\{$|C(\mathbf{a},\mathbf{d})(\tau)+C(\mathbf{b},\mathbf{c})(\tau)|\}_{\tau=-17}^{17}\\=( 0    0    0   0    0   4    0    4   8    0    12    0    4    4    4   4   12    4    4    4    12    4   4   0   12    0   0    4   8   4   0    0    0    0    0) $.
\end{example}
\begin{remark}
Although, the CZCPs generated from \textit{Lemma} \ref{lem4} and \cite[Th. 1]{czcp1} have the same length and ZCZ width, but these pairs are totally different. In fact, the CZCPs obtained from \textit{Lemma} \ref{lem4} is a complementary mate of CZCPs obtained from \cite[Th. 1]{czcp1}.
\end{remark}
In next theorem, construction of CZCSS of non-power-of-two length and large set size is provided, which has not been reported before.
\begin{theorem}\label{thm1}
For $d\in \{1,2\}$, let $\mathcal{F}_d^{\mathbf{t}},\mathcal{G}_d^{\mathbf{t}}:\{0,1\}^{m+n} \rightarrow \mathbb{Z}_q$ be defined by 
\begin{equation}\label{eqn19}
    \mathcal{F}_d^{\mathbf{t}}(\mathbf{x},\mathbf{y})= G_d+\frac{q}{2}\mathbf{t}\cdot \mathbf{y};~\mathcal{G}_d^{\mathbf{t}}(\mathbf{x}, \mathbf{y})=G_{d}+\frac{q}{2} x_{m-2}+\frac{q}{2} \mathbf{t} \cdot \mathbf{y},
\end{equation}
where $\mathbf{t}=\left(t_{0}, t_{1}, \ldots, t_{n-1}\right) \in \mathbb{Z}_{2}^{n}, \mathbf{x}=\left(x_{0}, x_{1}, \ldots, x_{m-1}\right)$, and $\mathbf{y}=\left(y_{0}, y_{1}, \ldots, y_{n-1}\right)$. Also, let $S_{\mathbf{t}}$ and $S_{\mathbf{t}}^{\prime}$ be two sets defined by
\begin{equation}\label{eqn21}
S_{\mathbf{t}}=\left\{\Psi_{2^{m-2}-1}\left(\mathcal{F}_d^{\mathbf{t}}(\mathbf{x}, \mathbf{y})\right): d \in \{1,2\}, \mathbf{y} \in \mathbb{Z}_{2}^{n}\right\},
\end{equation}
and
\begin{equation}\label{eqn22}
S_{\mathbf{t}}'=\left\{\Psi_{2^{m-2}-1}\left(\mathcal{G}_d^{\mathbf{t}}(\mathbf{x}, \mathbf{y})\right): d \in \{1,2\}, \mathbf{y} \in \mathbb{Z}_{2}^{n}\right\},
\end{equation}
for some fixed $\mathbf{t}$. Then $\left\{S_{\mathbf{t}}: \mathbf{t} \in \mathbb{Z}_{2}^{n}\right\} \cup\left\{S_{\mathbf{t}'}': \mathbf{t}' \in \mathbb{Z}_{2}^{n}\right\}$ forms a $\left(2^{n+1},2^{n+1},2^{m-1}+2,2^{\pi(m-3)}+1\right)$-CZCSS.
\end{theorem}
\begin{IEEEproof}
For proving the result, we use the notation that $i$-{th} sequence of the set $S_{\mathbf{t}}$ is denoted by  $S_{\mathbf{t},i}$ and  $S_{t, 2^{n+1}+1}=S_{t, 1}$. Also define the sets $\mathcal{U}_1=\{1,2,\hdots,2^{\pi(m-3)}+1\}$ and $\mathcal{U}_2=\{2^{m-1}-2^{\pi(m-3)}+1,2^{m-1}-2^{\pi(m-3)}+2,\hdots,2^{m-1}+1\}$.
For $\mathbf{t}\in \mathbb{Z}_2^n$, the property P1 of \textit{Definition} \ref{def4} is calculated as 
\begin{equation}\label{eqn23}
\begin{aligned}
C\left(S_{\mathbf{t}}, S_{\mathbf{t}}\right)(\tau) &\!=\!\sum_{\left.d \in \{1, 2\right\}}\! \sum_{\mathbf{y} \in \mathbb{Z}_{2}^{n}}\! C\!\left(\!G_{d}+\frac{q}{2} \mathbf{t} \cdot \mathbf{y}, G_{d}+\frac{q}{2} \mathbf{t}\cdot \mathbf{y}\right)(\tau) \\
&=A\left(G_{1}\right)(\tau)+A\left(G_{2}\right)(\tau) \cdot \sum_{\mathbf{y} \in \mathbb{Z}_{2}^{n}} \omega^{\frac{q}{2}\left(\mathbf{t}-\mathbf{t}\right) \cdot \mathbf{y}}\\
&=2^{n}\left(A\left(G_{1}\right)(\tau)+A\left(G_{2}\right)(\tau)\right)\\&=0~ \text{for}~ |\tau| \in \mathcal{U}_1 \cup \mathcal{U}_2 .
\end{aligned}
\end{equation}
The above equation becomes zero as $G_1$ and $G_2$ forms a CZCP. Since $G_1+\frac{q}{2}x_{m-2}$ and $G_2+\frac{q}{2}x_{m-2}$ is also a CZCP, similarly it can also be shown that $C\left(S'_{\mathbf{t}}, S'_{\mathbf{t}}\right)(\tau)=0$, for all $|\tau| \in \mathcal{U}_1 \cup \mathcal{U}_2 $. 
For proving the property P2 of \textit{Definition} \ref{def4}, consider the following cross-correlation
\begin{equation}\label{eqn24}
\begin{aligned}
    &\sum_{i=1}^{2^{n+1}} C\left(S_{\mathbf{t}, i}, S_{\mathbf{t}, i+1}\right)(\tau)\\=&
 \sum_{\mathbf{y}=0}^{2^{n}-2} C\left(G_{1}+\frac{q}{2} \mathbf{t} \cdot \mathbf{y}, G_{1}+\frac{q}{2} \mathbf{t} \cdot\mathbf{(y+1)}\right)(\tau) \\
&+C\left(G_{1}+\frac{q}{2} \mathbf{t} \cdot\left(\mathbf{2^{n}-1}\right), G_{2}+\frac{q}{2} \mathbf{t} \cdot \mathbf{0}\right)(\tau) \\
&+ \sum_{\mathbf{y}=0}^{2^{n}-2} C\left(G_{2}+\frac{q}{2} \mathbf{t} \cdot \mathbf{y}, G_{2}+\frac{q}{2} \mathbf{t} \cdot\mathbf{(y+1)}\right)(\tau)\\
&+C\left(G_{2}+\frac{q}{2} \mathbf{t} \cdot\left(\mathbf{2^{n}-1}\right), G_{1}+\frac{q}{2} \mathbf{t} \cdot \mathbf{0}\right)(\tau)\\
=&\sum_{\mathbf{y}=0}^{2^{n}-2}\left(\omega^{\frac{q} {2} \cdot\mathbf{t}}\left(A(G_1)(\tau)+A(G_2)(\tau)\right)\right)\\&+\omega^{\frac{q}{2}\cdot\mathbf{t(2^n-1)}}\left(C(G_1,G_2)(\tau)+C(G_2,G_1)(\tau)\right)\\=&0,~ \text{for all}~ |\tau| \in \mathcal{U}_2.~ \text{(as $G_1,G_2$ is a CZCP)}
\end{aligned}
\end{equation}
Similarly, it can also be shown that 
    $\sum_{i=1}^{2^{n+1}} C\left(S_{t, i}^{\prime}, S_{t, i+1}^{\prime}\right)(\tau)$=$0$, for all $|\tau| \in \mathcal{U}_{2}$. So, the property P2 of \textit{Definition} \ref{def4} is proved. For proving the property P3 of \textit{Definition} \ref{def4}, the following different cases need to be proved.
    For $\mathbf{t} \neq \mathbf{t}'\in \mathbb{Z}_2^n$, we need to prove $C(S_{\mathbf{t}},S_{\mathbf{t}'})=0$,   $C(S'_{\mathbf{t}},S'_{\mathbf{t}'})=0$, for all $|\tau| \in \{0\}\cup\mathcal{U}_1 \cup \mathcal{U}_2$, and for any $\mathbf{t}, \mathbf{t}'\in \mathbb{Z}_2^n$, for all $|\tau|  \in \{0\} \cup \mathcal{U}_1 \cup \mathcal{U}_2$, $C(S_{\mathbf{t}},S'_{\mathbf{t}'})=0$. Consider the cross-correlation between sets $\mathbf{S}_t$ and $\mathbf{S}_{t'}$
\begin{equation}\label{eqn25}
\begin{aligned}
&C\left(S_{\mathbf{t}}, S_{\mathbf{t}'}\right)(\tau)\\&=\sum_{\left.d \in \{1, 2\right\}} \sum_{\mathbf{y} \in \mathbb{Z}_{2}^{n}} C\left(G_{d}+\frac{q}{2} \mathbf{t} \cdot \mathbf{y}, G_{d}+\frac{q}{2} \mathbf{t}'\cdot \mathbf{y}\right)(\tau) \\
&=\left(A\left(G_{1}\right)(\tau)+A\left(G_{2}\right)(\tau)\right) \cdot \sum_{\mathbf{y} \in \mathbb{Z}_{2}^{n}} \omega^{\frac{q}{2}\left(\mathbf{t}-\mathbf{t}'\right) \cdot \mathbf{y}}.
\end{aligned}
\end{equation}
Since, for $\mathbf{t} \neq \mathbf{t}'$, $\sum_{\mathbf{y} \in\mathbb{Z}_{2}^{n}} \omega^{q / 2\left(\mathbf{t}-\mathbf{t}'\right) \cdot \mathbf{y}} =0$. So, from (\ref{eqn25}) $C\left(S_{\mathbf{t}}, S_{\mathbf{t}'}\right)(\tau)=0$, for all $\tau$. Similarly, $C\left(S'_{\mathbf{t}}, S'_{\mathbf{t}'}\right)(\tau)=0$, for all $\tau$. Now for any $\mathbf{t}, \mathbf{t}'\in \mathbb{Z}_2^n$, consider the cross-correlation between the sets
\begin{equation}\label{eqn26}
\begin{aligned}
&C\left(S_{\mathbf{t}}, S'_{\mathbf{t}'}\right)(\tau)\\ &=\sum_{\left.d \in \{1, 2\right\}} \sum_{\mathbf{y} \in \mathbb{Z}_{2}^{n}} C\left(G_{d}+\frac{q}{2} \mathbf{t} \cdot \mathbf{y}, G_{d}+\frac{q}{2}x_{m-2}+\frac{q}{2} \mathbf{t}'\cdot \mathbf{y}\right)(\tau) \\
&=\left(C\left(G_{1},G_1+\frac{q}{2}x_{m-2}\right)(\tau)+C\left(G_{2},G_2+\frac{q}{2}x_{m-2}\right)(\tau)\right) \\&\cdot \sum_{\mathbf{y} \in \mathbb{Z}_{2}^{n}} \omega^{\frac{q}{2}\left(\mathbf{t}-\mathbf{t}'\right) \cdot \mathbf{y}}\\
&=0,~ \text{for all}~ \tau . ~\text{(using \textit{Lemma} \ref{lem4})}
\end{aligned}
\end{equation}
Now, the two cases for the property P4 of \textit{Definition} \ref{def4} namely, for any $\mathbf{t}\neq \mathbf{t}'\in \mathbb{Z}_2^n$, $\sum_{i=1}^{2^{n+1}}C\left(S_{\mathbf{t}, i}, S_{\mathbf{t}', i+1}\right)(\tau)=0$  and $\sum_{i=1}^{2^{n+1}}C\left(S'_{\mathbf{t}, i}, S'_{\mathbf{t}', i+1}\right)(\tau)=0$ for all $|\tau| \in \mathcal{U}_2$  can also be proved in a similar way as the property P3 is proved. Here, we prove the remaining case, i.e.,
 for any $\mathbf{t}, \mathbf{t}'\in \mathbb{Z}_2^n$, consider the cross-correlation term
\begin{equation}\label{eqn28}
\begin{aligned}
    &\sum_{i=1}^{2^{n+1}}C\left(S'_{\mathbf{t}, i}, S_{\mathbf{t}', i+1}\right)(\tau) \\
= & \sum_{\mathbf{y}=0}^{2^{n}-2} C\left(G_{1}+\frac{q}{2} \mathbf{t} \cdot \mathbf{y}, G_{1}+\frac{q}{2}x_{m-2}+\frac{q}{2} \mathbf{t}' \cdot\mathbf{(y+1)}\right)(\tau) \\
&+C\left(G_{1}+\frac{q}{2} \mathbf{t} \cdot\left(\mathbf{2^{n}-1}\right), G_{2}+\frac{q}{2}x_{m-2}+\frac{q}{2} \mathbf{t}' \cdot \mathbf{0}\right)(\tau) \\
&+ \sum_{\mathbf{y}=0}^{2^{n}-2} C\left(G_{2}+\frac{q}{2} \mathbf{t} \cdot \mathbf{y}, G_{2}+\frac{q}{2}x_{m-2}+\frac{q}{2} \mathbf{t}' \cdot\mathbf{(y+1)}\right)(\tau)\\
&+C\left(G_{2}+\frac{q}{2} \mathbf{t} \cdot\left(\mathbf{2^{n}-1}\right), G_{1}+\frac{q}{2}x_{m-2}+\frac{q}{2} \mathbf{t}' \cdot \mathbf{0}\right)(\tau)\\
=&\left(C(G_1,G_1+\frac{q}{2}x_{m-2})(\tau)+C(G_2,G_2+\frac{q}{2}x_{m-2})(\tau)\right)\\&
\cdot\sum_{\mathbf{y}=0}^{2^{n}-2}\omega^{\frac{q}{2} \cdot(\mathbf{t}\mathbf{y}-\mathbf{t}'(\mathbf{y+1}))}\\&
+\left(C(G_1,G_2+\frac{q}{2}x_{m-2})(\tau)+C(G_2,G_1+\frac{q}{2}x_{m-2})(\tau)\right)\\& \cdot \omega^{\frac{q}{2}  \cdot\mathbf{t(2^n-1)}}\\=&0,~ \text{for all}~ |\tau| \in \mathcal{U}_2.
\end{aligned}
\end{equation}
The sum in (\ref{eqn28}) is zero from \textit{Lemma} \ref{lem4}.
\end{IEEEproof}
\begin{example}\label{ex2}
Let $n=2$ and all other parameters and functions are as defined in \textit{Example} \ref{ex1}. Then for $\mathbf{t}=(t_0,t_1) \in \{0,1\}^2$, and $\mathbf{y}=(y_0,y_1) \in \{0,1\}^2$, consider the sets
\begin{equation}\label{eqn29}
    S_{\mathbf{t}}=\left\{\Psi_7\left(G_d+\frac{q}{2}\mathbf{t}\cdot\mathbf{y})\right): d \in \{1,2\}, \mathbf{y} \in \mathbb{Z}_{2}^{2}\right\},
\end{equation}
and
\begin{equation}\label{eqn30}
    S'_{\mathbf{t}}=\left\{\Psi_7\left(G_d+\frac{q}{2}x_{m-2}+\frac{q}{2}\mathbf{t}\cdot\mathbf{y})\right): d \in \{1,2\}, \mathbf{y} \in \mathbb{Z}_{2}^{2}\right\}.
\end{equation}
Then from \textit{Theorem} \ref{thm1}  $\left\{S_{\mathbf{t}}: \mathbf{t} \in \mathbb{Z}_{2}^{2}\right\} \cup\left\{S_{\mathbf{t}'}': \mathbf{t}' \in \mathbb{Z}_{2}^{2}\right\}$ forms a $\left(8,8,18,5\right)$-CZCSS.
\end{example}
\begin{remark}\label{remark2}
The ZCZ width of the proposed CZCSS is maximum when $\pi(m-3)=m-3$.
\end{remark}
\begin{remark} In \cite{czcs} several concatenation based constructions of $(4,N,Z)$-CZCS is provided. Every CZCSS is a collection of CZCS, hence for the first time \textit{Theorem} \ref{thm1} also facilitate the GBF based construction of $(2^{n+1},2^{m-1}+2,2^{\pi(m-3)}+1)$-CZCSs.
\end{remark}
\begin{remark}
Since every CZCSS is also a SZCCS, so \textit{Theorem} \ref{thm1} also generates $(2^{n+1},2^{n+1},2^{m-1}+2,2^{\pi(m-3)}+1)$-SZCCS. The available construction of SZCCS in \cite{szccs1}, is limited to maximum set size $8$, this is for the first time that SZCCS with large set size is constructed.
\end{remark}
\section{Conclusion}
In this letter, we have proposed a direct construction of CZCSS of non-power-of-two length and large set size based on GBF, which has not been reported before. The set size of the proposed CZCSS is equal to the number of constituent sequences. The proposed construction also generates non-power-of-two length CZCSs containing a large number of sequences and non-power-of-two length SZCCS with a large set size. 
\bibliographystyle{IEEEtran}
\bibliography{reference}

\begin{thebibliography}{10}
\providecommand{\url}[1]{#1}
\csname url@samestyle\endcsname
\providecommand{\newblock}{\relax}
\providecommand{\bibinfo}[2]{#2}
\providecommand{\BIBentrySTDinterwordspacing}{\spaceskip=0pt\relax}
\providecommand{\BIBentryALTinterwordstretchfactor}{4}
\providecommand{\BIBentryALTinterwordspacing}{\spaceskip=\fontdimen2\font plus
\BIBentryALTinterwordstretchfactor\fontdimen3\font minus
  \fontdimen4\font\relax}
\providecommand{\BIBforeignlanguage}[2]{{%
\expandafter\ifx\csname l@#1\endcsname\relax
\typeout{** WARNING: IEEEtran.bst: No hyphenation pattern has been}%
\typeout{** loaded for the language `#1'. Using the pattern for}%
\typeout{** the default language instead.}%
\else
\language=\csname l@#1\endcsname
\fi
#2}}
\providecommand{\BIBdecl}{\relax}
\BIBdecl

\bibitem{fan2007}
P.~Fan, W.~Yuan, and Y.~Tu, ``Z-complementary binary sequences,'' \emph{IEEE
  Signal Process. Lett.}, vol.~14, no.~8, pp. 509--512, Aug. 2007.

\bibitem{zcp1}
C.-Y. Pai, S.-W. Wu, and C.-Y. Chen, ``Z-complementary pairs with flexible
  lengths from generalized {B}oolean functions,'' \emph{IEEE Commun. Lett.},
  vol.~24, no.~6, pp. 1183--1187, 2020.

\bibitem{zcp2}
C.-Y. Chen, ``A novel construction of {Z}-complementary pairs based on
  generalized {B}oolean functions,'' \emph{IEEE Signal Process. Lett.},
  vol.~24, no.~7, pp. 987--990, 2017.

\bibitem{zcp3}
A.~R. Adhikary, P.~Sarkar, and S.~Majhi, ``A direct construction of $q$-ary
  even length {Z}-complementary pairs using generalized {B}oolean functions,''
  \emph{IEEE Signal Process. Lett.}, vol.~27, pp. 146--150, 2020.

\bibitem{zcp4}
A.~R. Adhikary, S.~Majhi, Z.~Liu, and Y.~L. Guan, ``New sets of optimal
  odd-length binary {Z}-complementary pairs,'' \emph{IEEE Trans. Inf. Theory},
  vol.~66, no.~1, pp. 669--678, 2020.

\bibitem{praveen1}
P.~Kumar, P.~Sarkar, S.~Majhi, and S.~Paul, ``A direct construction of even
  length {ZCPs} with large {ZCZ} ratio,'' \emph{Cryptogr. Commun.}, 2022.

\bibitem{czcp2}
Z.~Liu, P.~Yang, Y.~L. Guan, and P.~Xiao, ``Cross {Z}-complementary pairs for
  optimal training in spatial modulation over frequency selective channels,''
  \emph{IEEE Trans. Signal Process.}, vol.~68, pp. 1529--1543, 2020.

\bibitem{czcp1}
A.~R. Adhikary, Z.~Zhou, Y.~Yang, and P.~Fan, ``Constructions of cross
  {Z}-complementary pairs with new lengths,'' \emph{IEEE Trans. Signal
  Process.}, vol.~68, pp. 4700--4712, 2020.

\bibitem{czcp3}
M.~Yang, S.~Tian, N.~Li, and A.~R. Adhikary, ``New sets of quadriphase cross
  {Z}-complementary pairs for preamble design in spatial modulation,''
  \emph{IEEE Signal Process. Lett.}, vol.~28, pp. 1240--1244, 2021.

\bibitem{czcp4}
Z.-M. Huang, C.-Y. Pai, and C.-Y. Chen, ``Binary cross {Z}-complementary pairs
  with flexible lengths from {B}oolean functions,'' \emph{IEEE Commun. Lett.},
  vol.~25, no.~4, pp. 1057--1061, 2021.

\bibitem{czcp5}
C.~Fan, D.~Zhang, and A.~R. Adhikary, ``New sets of binary cross
  {Z}-complementary sequence pairs,'' \emph{IEEE Commun. Lett.}, vol.~24,
  no.~8, pp. 1616--1620, 2020.

\bibitem{czcp6}
F.~Zeng, X.~He, Z.~Zhang, and L.~Yan, ``Quadriphase cross {Z}-complementary
  pairs for pilot sequence design in spatial modulation systems,'' \emph{IEEE
  Signal Process. Lett.}, vol.~29, pp. 508--512, 2022.

\bibitem{czcp7}
Z.~Liu, P.~Yang, Y.~L. Guan, and P.~Xiao, ``Cross {Z}-complementary pairs
  ({CZCPs}) for optimal training in broadband spatial modulation systems,'' in
  \emph{2020 IEEE Int. Symp. Inf. Theory ({ISIT})}, 2020, pp. 2926--2930.

\bibitem{czcs}
Z.-M. Huang, C.-Y. Pai, and C.-Y. Chen, ``Cross {Z}-complementary sets for
  training design in spatial modulation,'' \emph{IEEE Trans. Commun.}, pp.
  1--1, 2022.

\bibitem{zccs1}
L.~Feng, P.~Fan, X.~Tang, and K.-k. Loo, ``Generalized pairwise
  {Z}-complementary codes,'' \emph{IEEE Signal Process. Lett.}, vol.~15, pp.
  377--380, 2008.

\bibitem{pmz}
P.~Sarkar, S.~Majhi, and Z.~Liu, ``Optimal {Z}-complementary code set from
  generalized {R}eed-{M}uller codes,'' \emph{IEEE Trans. Commun.}, vol.~67,
  no.~3, pp. 1783--1796, 2019.

\bibitem{zccs2}
------, ``Pseudo-{B}oolean functions for optimal {Z}-complementary code sets
  with flexible lengths,'' \emph{IEEE Signal Process. Lett.}, vol.~28, pp.
  1350--1354, 2021.

\bibitem{zccs3}
P.~Sarkar, A.~Roy, and S.~Majhi, ``Construction of {Z}-complementary code sets
  with non-power-of-two lengths based on generalized {B}oolean functions,''
  \emph{IEEE Commun. Lett.}, vol.~24, no.~8, pp. 1607--1611, 2020.

\bibitem{gobinda}
G.~Ghosh, S.~Majhi, P.~Sarkar, and A.~K. Upadhaya, ``Direct construction of
  optimal {Z}-complementary code sets with even lengths by using generalized
  {B}oolean functions,'' \emph{IEEE Signal Process. Lett.}, vol.~29, pp.
  872--876, 2022.

\bibitem{szccs1}
\BIBentryALTinterwordspacing
Y.~Zhou, Z.~Zhou, Z.~Liu, Y.~Yang, P.~Yang, and P.~Fan, ``Symmetrical
  {Z}-complementary code sets ({SZCCSs}) for optimal training in generalized
  spatial modulation,'' 2020. [Online]. Available:
  \url{https://arxiv.org/abs/2010.11372}
\BIBentrySTDinterwordspacing

\bibitem{wuyu}
S.-W. Wu and C.-Y. Chen, ``Optimal {Z}-complementary sequence sets with good
  peak-to-average power-ratio property,'' \emph{IEEE Signal Process. Lett.},
  vol.~25, no.~10, pp. 1500--1504, 2018.

\end{thebibliography}
\end{document}